\documentclass[10pt,conference]{IEEEtran}
\IEEEoverridecommandlockouts
\def\BibTeX{{\rm B\kern-.05em{\sc i\kern-.025em b}\kern-.08em
    T\kern-.1667em\lower.7ex\hbox{E}\kern-.125emX}}
\usepackage{listings}
\usepackage{color}
\usepackage{xspace}
\usepackage{xcolor}
\usepackage{graphicx}
\usepackage{stfloats}
\usepackage{enumitem}
\usepackage{amsmath}
\usepackage{fancybox}
\usepackage[ruled,lined,linesnumbered,vlined,algo2e]{algorithm2e}
\usepackage{soul}
\usepackage{url}
\usepackage{calc}
\usepackage[many]{tcolorbox}
\newcommand{\distance}{2pt}
\usepackage{titlesec}
\usepackage{booktabs}
\usepackage{setspace}

\titlespacing*{\section}{0pt}{0.5\baselineskip}{0.3\baselineskip}
\titlespacing*{\subsection}{0pt}{0.5\baselineskip}{0.2\baselineskip}

\usepackage[font=small,skip=2pt]{caption}

\usepackage{microtype}
\setlist[itemize]{noitemsep, topsep=0pt}
\definecolor{OliveGreen}{rgb}{0,0.6,0}
\newcommand{\ly}[1][\textcolor{black}]{#1}

\usepackage{cleveref}

\newcommand{\depv}{\text{dependency version}\xspace}
\newcommand{\tool}{\text{Ranger}\xspace}
\newcommand{\soft}{\text{SoftVer}\xspace}
\newcommand{\softs}{\text{SoftVers}\xspace}
\newcommand{\searchalg}{\text{ALSearch}\xspace}

\usepackage{cite}

\author{
\IEEEauthorblockN{Lyuye Zhang\IEEEauthorrefmark{5}\IEEEauthorrefmark{1},
Chengwei Liu\IEEEauthorrefmark{1}\IEEEauthorrefmark{4},
Sen Chen\IEEEauthorrefmark{2},
Zhengzi Xu\IEEEauthorrefmark{1},
Lingling Fan\IEEEauthorrefmark{3},
Lida Zhao\IEEEauthorrefmark{1},
Yiran Zhang\IEEEauthorrefmark{1},
Yang Liu\IEEEauthorrefmark{1}
}
\text{\small zh0004ye@e.ntu.edu.sg, 
chengwei001@e.ntu.edu.sg}
\IEEEauthorblockA{\IEEEauthorrefmark{5}Continental-NTU Corporate Lab, Nanyang Technological University, Singapore}
\thanks{\IEEEauthorrefmark{4} Chengwei Liu is the corresponding author.}
\IEEEauthorblockA{\IEEEauthorrefmark{1}School of Computer Science and Engineering, Nanyang Technological University, Singapore}

\IEEEauthorblockA{\IEEEauthorrefmark{2}College of Intelligence and Computing, Tianjin University, China}
\IEEEauthorblockA{\IEEEauthorrefmark{3}College of Cyber Science, Nankai University, China}

}

\title{Mitigating Persistence of Open-Source Vulnerabilities in Maven Ecosystem}

\usepackage[font=small,skip=2pt]{caption}
\begin{document}
\abovedisplayskip=0pt
\abovedisplayshortskip=0pt
\belowdisplayskip=0pt
\belowdisplayshortskip=0pt
\maketitle
\thispagestyle{plain}
\pagestyle{plain}

\begin{abstract}
Vulnerabilities from third-party libraries (TPLs) have been unveiled to threaten the Maven ecosystem in the long term. Despite patches being released promptly after vulnerabilities are disclosed, the libraries and applications in the community still use the vulnerable versions, which makes the vulnerabilities persistent in the Maven ecosystem (e.g., the notorious Log4Shell still greatly influences the Maven ecosystem nowadays from 2021).  
Both academic and industrial researchers have proposed user-oriented standards and solutions to address vulnerabilities, while such solutions fail to tackle the ecosystem-wide persistent vulnerabilities because it requires a collective effort from the community to timely adopt patches without introducing breaking issues.

To seek an ecosystem-wide solution, we first carried out an empirical study to examine the prevalence of persistent vulnerabilities in the Maven ecosystem. Then, we identified affected libraries for alerts by implementing an algorithm monitoring downstream dependents of vulnerabilities based on an up-to-date dependency graph. Based on them, we further quantitatively revealed that patches blocked by upstream libraries caused the persistence of vulnerabilities. After reviewing the drawbacks of existing countermeasures, to address them, we proposed a solution for \underline{range r}estoration (\tool) to automatically restore the compatible and secure version ranges of dependencies for downstream dependents. The automatic restoration requires no manual effort from the community, and the code-centric compatibility assurance ensures smooth upgrades to patched versions. Moreover, \tool along with the ecosystem monitoring can timely alert developers of blocking libraries and suggest flexible version ranges to rapidly unblock patch versions. By evaluation, \tool could restore $75.64\%$ of ranges which automatically remediated $90.32\%$ of vulnerable downstream projects. 
\end{abstract}


\section{Introduction}
\label{sec:introduction}

The vulnerabilities present in widely used TPLs have garnered significant attention from communities. 
\textit{log4j-core}, serving as a fundamental library, swiftly responded by releasing patch updates after the exploitation of Log4Shell~\cite{log4jvulrce}. Downstream users have taken prompt action to adopt these patch updates, as reported by Google~\cite{osvinsight}. 
Despite a year's worth of advancements, Log4Shell continues to impact numerous downstream applications and persist within the Maven ecosystem, as reported by many reports and news~\cite{log4jnews1,log4jnews2,log4jnews3,log4jnews4,log4jnews5}. Given that over 2,000 vulnerabilities from Maven libraries have been disclosed by the National Vulnerability Database (NVD)~\cite{nvd}, it is possible that numerous other vulnerabilities persist and pose a threat to the Maven ecosystem.

Aiming for this urgent threat, many researchers \cite{wu2023understanding,li2021pdgraph,pashchenko2020qualitative,mir2023effect,soto2019emergence,benelallam2019maven,du2018refining} studied the vulnerability impact within the Maven ecosystem
and substantiated vulnerabilities have extensively proliferated in downstream libraries.
\ly{A few of them have recognized the persistence of vulnerabilities over time and provided insights into potential solutions~\cite{wu2023understanding, li2021pdgraph, pashchenko2020qualitative, wang2023plumber}.}
Wu et al.~\cite{wu2023understanding, imtiaz2023open} revealed that the reachable vulnerabilities are more likely to be addressed. 
Developers' reluctance to upgrade vulnerable dependencies due to potential breaking changes has been highlighted by Pashchenko et al.~\cite{pashchenko2020qualitative} who also discovered that developers prioritize handling vulnerabilities in direct dependencies rather than transitive ones~\cite{pashchenko2018vulnerable}. 
Moreover, Industrial standards have been proposed to promote remediation, such that OpenSSF~\cite{openssf} proposed the best practice guidance~\cite{ossfwgbe} and a tool, Scorecard~\cite{scorecard}, for developers on managing vulnerabilities in dependencies. Plumber~\cite{wang2023plumber} aims for persistent vulnerabilities in Node Package Manager (NPM) with limited applicability to Maven due to the rare usage of version ranges in Maven. \ly{However, because these solutions are either user-oriented aiming for individual projects or inapplicable for Maven, not all stakeholders in the ecosystem would be benefited, which barely promotes the ecosystem-wide mitigation of persistent vulnerabilities due following issues:}

\textbf{Issue 1: The lack of collective awareness.} Effectively mitigating persistent vulnerabilities requires collective efforts from the community, particularly from developers of widely-used libraries, rather than just a few individuals. Hence, the ability to accurately locate influential libraries and swiftly arouse awareness of relevant developers is missing yet required. 

\textbf{Issue 2: The overreliance on human practices.} 
Although developers may be aware of the negative consequences of vulnerabilities, they require a solid understanding of software security to effectively remediate them. Even if they possess the necessary skills, remediation practices such as upgrading, backporting, and migration are often time-consuming and require significant manual effort. As such, relying solely on human practices to eliminate vulnerabilities within the software ecosystem is not realistic.

\textbf{Issue 3: The backward-incompatibility of dependencies.} Maven libraries are known to have version releases violating Semantic Versioning (SemVer)~\cite{raemaekers2014semantic,raemaekers2017semantic,lam2020putting,decan2019package}. Consequently, numerous breaking changes across upgrades may lurk within the ecosystem. To maintain stability, many developers prefer to define dependencies using single versions~\cite{dietrich2019dependency}, rather than flexible version ranges, even though Maven allows for the latter. It further complicates the mitigation of vulnerabilities, as automatic security upgrades are not widely applicable, in contrast to the NPM ecosystem~\cite{liu2022demystifying}.

\begin{figure}[!t]
  \centering \includegraphics[width=1\linewidth]{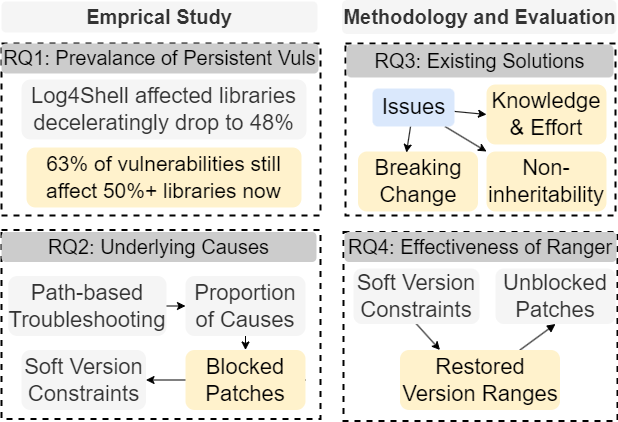}
  \caption{Overview} 
  \label{fig:overview}
\end{figure}

To address the issues outlined above, we did the followings:
\begin{itemize}[leftmargin=5pt]
\item For \textbf{Issue 1}, as illustrated in Figure~\ref{fig:overview}, we first studied the prevalence of persistent vulnerabilities within the Maven ecosystem and identified affected libraries for alerts.  
Specifically, we implemented an algorithm based on a dependency vulnerability graph we constructed to recursively identify downstream vulnerable dependents. 
Based on these, the impact of persistent vulnerabilities is uncovered regarding time span and affected libraries\footnote{Affected libraries refer to the libraries that have the vulnerabilities in their direct or transitive deployable dependencies} in the Maven ecosystem (RQ1). Our study revealed that, upon disclosure, approximately 82.22\% of vulnerabilities within the Maven ecosystem remain unresolved in over 50\% of the downstream libraries. As of the date of data collection, $58.73\%$ of these vulnerabilities still impacted more than $50\%$ of downstream libraries. Furthermore, it is revealed that persistent vulnerabilities are caused by blocked fixes by downstream libraries, and blocking libraries can be accurately located with our algorithm (RQ2). 
\item For \textbf{Issue 2}, we explored the effectiveness of existing countermeasures in remediating persistent vulnerabilities in the Maven ecosystem. However, we found that these workarounds either required extensive manual effort or were susceptible to breaking changes, highlighting the need for an automatic, scalable and compatibility assurable solution (RQ3). 
\item For \textbf{Issue 3}, we propose a solution for \underline{range r}estoration (\tool) for both clients and the Maven ecosystem to automatically restore compatible and secure version ranges for vulnerable libraries and dependents. \tool checks all types of code-centric compatibility with state-of-the-art tools to exclude breaking versions and employs unit tests for validation. With compatible version ranges, patched versions of vulnerable libraries and dependencies could be automatically resolved for downstream users without human intervention. \tool also continues to mitigate persistent vulnerabilities in the ecosystem by monitoring blocking libraries and providing range suggestions to relevant developers to arouse community awareness. In the evaluation, \tool could restore $3,109$ ($75.64\%$) ranges which automatically remediated $10, 678$ ($90.32\%$) vulnerable downstream projects (RQ4). 
\end{itemize}

The contributions we made are as follows:
\begin{itemize}[leftmargin=9pt]
    \item We developed \tool to restore compatible and secure version ranges which could automatically mitigate the persistent vulnerabilities in the Maven ecosystem.
    \item We conducted an empirical study to substantiate the persistence of vulnerabilities and quantitatively revealed their underlying cause and the effectiveness of countermeasures. 
    \item We implemented a monitoring system based on an up-to-date dependency graph and a search algorithm to locate the libraries that block vulnerability fixes and suggest remediation for relevant developers and downstream users.
\end{itemize}

\section{Preparation for Empirical Study}
\label{sec:method}
To commence our study, we first briefly introduce the concept of SemVer used in Maven. Then, we constructed a dependency graph using data sourced from both the Maven Central Repository (MCR) and NVD. Based on the dependency graph, we developed a searching algorithm (\searchalg) to facilitate tracking of affected libraries throughout the course of our study.

\subsection{Background of SemVer in Maven}
Within the Maven ecosystem, most version numbers adhere to the SemVer standard~\cite{semver}. This standard consists of three digits: \textit{Major}, \textit{Minor}, and \textit{Patch}.
Major upgrades, which change the \textit{Major} digit, are the only type of upgrade that allow for incompatible changes.
Version ranges~\cite{mavenranges} supported by Maven rely on SemVer. However, $99.21\%$ of dependency version specifications in Maven are single versions which are called Soft Version Constraints~\cite{mavensoft} (\soft). The \soft stipulates the preferred version for a dependency so that Maven mostly resolves the preferred versions for the dependencies~\cite{dietrich2019dependency}.

\subsection{Infrastructure of Study}
\subsubsection{Dependency Graph for Maven}
\label{sec:graph}
A dependency graph was constructed, including vulnerabilities, as an infrastructure for the empirical analysis.
As of 01 Apr 2023, MCR contained $541,753$ libraries and $11,859,883$ versions, both of which were extracted from the MCR index~\cite{mvnrepo} and added to the dependency graph as \textit{Library} and \textit{Version} vertices. $82,708,563$ dependency edges from \textit{Version} to \textit{Version} were extracted from the Project Object Model (POM) files including properties specifically designed to regulate the dependency resolution.
We used the approximately over 2k Common Vulnerabilities and Exposures (CVE) for Maven libraries at NVD as vulnerability data. Due to the absence of well-formatted mappings between vulnerabilities and versions, $1,861$ vulnerabilities and their mappings were collected after cross-checking multiple sources from Github Advisory~\cite{githubadvisory}, Google Open-Source Database~\cite{googledataset}, and Snyk Vulnerability Database~\cite{snykvuldb}, which are available on our website~\cite{dataset}.

\subsubsection{Search Algorithm}
\label{sec:deptree}
We developed a precise Affected Library Searching Algorithm (\searchalg) that leverages the Maven dependency resolution rules to accurately track dependents of vulnerabilities based on the dependency graph. Unlike the forward resolution approach used by Maven to resolve dependencies from the root to leaf vertices, \searchalg was designed to facilitate backward tracking from vulnerability vertices to dependent vertices. As \searchalg is tailored for backward tracking, its rules have been adapted accordingly, and are outlined below:

\textbf{Scope} is a feature to limit the transitivity of a dependency. Out of the six scopes, only \textit{compile} and \textit{runtime} are inheritable and tracked by \searchalg.
{\textbf{Optional} dependencies are not transitive, and thus should not be tracked} for dependents with $\geq 2$ depth.
\textbf{Exclusions} are used to exclude certain versions of transitive dependencies. All transitive dependencies under the \textit{exclusions} are excluded. 
Hence, if the libraries with vulnerabilities are excluded by any dependent, dependents should not be tracked.
\textbf{Multiple versions selection}: If a library is used with different versions in a dependency tree, Maven would prioritize the version specified first during a Breadth-First Search (BFS) resolution from direct to transitive dependencies. \searchalg considers a target library affected only if the vulnerable versions of the affected library are closer to the target library than the non-vulnerable versions.

Incorporating the above rules, for each vulnerability, \searchalg iterates over downstream libraries in a BFS manner. During each iteration, it includes two procedures to track a dependent and validate the tracked target respectively: 
\begin{itemize}[leftmargin=5pt]
    \item \textbf{Dependents tracking}: Check if the \textit{Version} vertex has  consecutive dependency edges pointing to any vulnerable version of a library affected by a \textit{Vulnerability} vertex. If yes, check if the properties on dependency edges adhere to the aforementioned rules. If yes, proceed to the next procedure.
    \item \textbf{Dependencies validation}:
    Resolve dependencies of the target \textit{Version} vertex following normal Maven dependency resolution rules in a reversed direction until the version of the vulnerable library is resolved. If the resolved version is vulnerable, the target \textit{Version} vertex is considered an affected version.
\end{itemize}
 
After the iteration, the affected \textit{Version} vertex is stored with the publishing date and depth.
To boost performance, the maximum depth of the call chain is initially set to 10, based on research indicating that the semantics decline after 10 successive calls~\cite{schroter2010stack}. Our study later also confirms that there are significantly fewer affected libraries beyond a depth of 9.
We verified \searchalg by randomly selecting $1,000$ affected library versions and retrieving dependency trees of them using the \textit{mvn deptree} command. If the library did depend on a vulnerable dependency, the library was considered affected. Only $12$ ($1.20\%$) libraries were false positives, mainly due to different OS requirements or incomplete data in MCR (discussed in Threat of Validity Section~\ref{sec:threat}).

     
        
        
        
            
                    

\section{Empirical Study}
\label{sec:evaluation}
To quantitatively assess the prevalence and underlying cause of persistent vulnerabilities in the Maven ecosystem, we conducted an empirical study to answer the following research questions:

\noindent $\bullet$ \textbf{RQ1}: \textbf{\textit{How prevalent are persistent vulnerabilities in the Maven ecosystem?}}
The impact of vulnerabilities over time is evaluated regarding the distribution of time spans and counts of affected libraries to demonstrate persistent vulnerabilities.

\noindent $\bullet$ \textbf{RQ2}: \textbf{\textit{What are the causes of persistent vulnerabilities?}}
We quantitatively uncovered the underlying factors by categorizing and analyzing 6 cases to identify the primary cause.

\noindent\textbf{Dataset:} the primary dataset is the dependency graph in Section~\ref{sec:graph}. To investigate the prevalence of Log4Shell in real-world projects, besides data from MCR, an additional dataset was created by cloning Java repositories on GitHub managed by Maven (with POM files) and filtering out those with fewer than 20 stars to ensure their popularity. As of April 1, 2023, a total of 13,638 repositories were collected, and dependency trees were extracted using the Maven command \textit{mvn deptree}. The dependency trees of 9,220 repositories were successfully extracted.

\subsection{RQ1: Analysis of persistent Vulnerabilities}
The impact of persistent vulnerabilities on downstream affected libraries is demonstrated by their long-tail prevalence. We used Log4Shell as an example to showcase the metrics we used and then evaluated all vulnerabilities to demonstrate the persistence.
\subsubsection{Log4Shell Analysis} 
First, we retrieved the affected library and version vertices associated with release dates with \searchalg. 
Because usually, the latest version of a library is the release currently maintained by the developers, a library is considered affected if the latest version depends on vulnerable \textit{log4j-core}.
The downstream libraries were categorized into 3 categories (1) \textbf{Affected}: The downstream library's latest version is affected. The proportion of these libraries is denoted as $P_{vul}$. (2) \textbf{Patched}: The library's latest version is not affected, but at least one of its previous versions was affected. The proportion of them is called $P_{patch}$. (3) \textbf{Removed}: The older versions of the library were affected, and the latest version does not depend on \textit{log4j-core} anymore.

\begin{figure}[!t]
  \centering \includegraphics[width=1\linewidth]{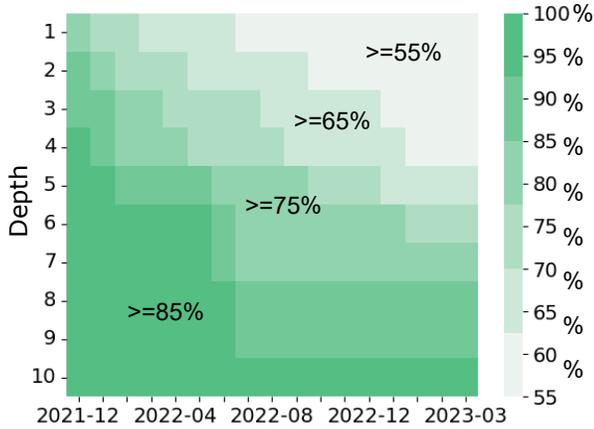}
  \caption{Heatmap of Proportion of Affected Libraries by Log4Shell} 
  \label{fig:heat}
\end{figure}

The $P_{vul}$ of Log4Shell over time is demonstrated in the heat map Figure~\ref{fig:heat}. In the heatmap, the x-axis refers to the timeline from the publishing date at NVD to 01 Apr 2023 based on months, while the y-axis refers to the depth. It is shown that $P_{vul}$ at depth $1$ decays faster than at other depths. It is because those libraries serve as first-level dependents, which would be quickly aware of the vulnerable versions of \textit{log4j-core} in their dependencies. With the depth increasing, the downstream libraries are less likely to be aware of the transitive vulnerability and less likely to execute the vulnerable code of \textit{log4j-core}. Thus, the decaying rate decreases as the depth goes deep.

\begin{figure}[!t]
  \centering \includegraphics[width=1\linewidth]{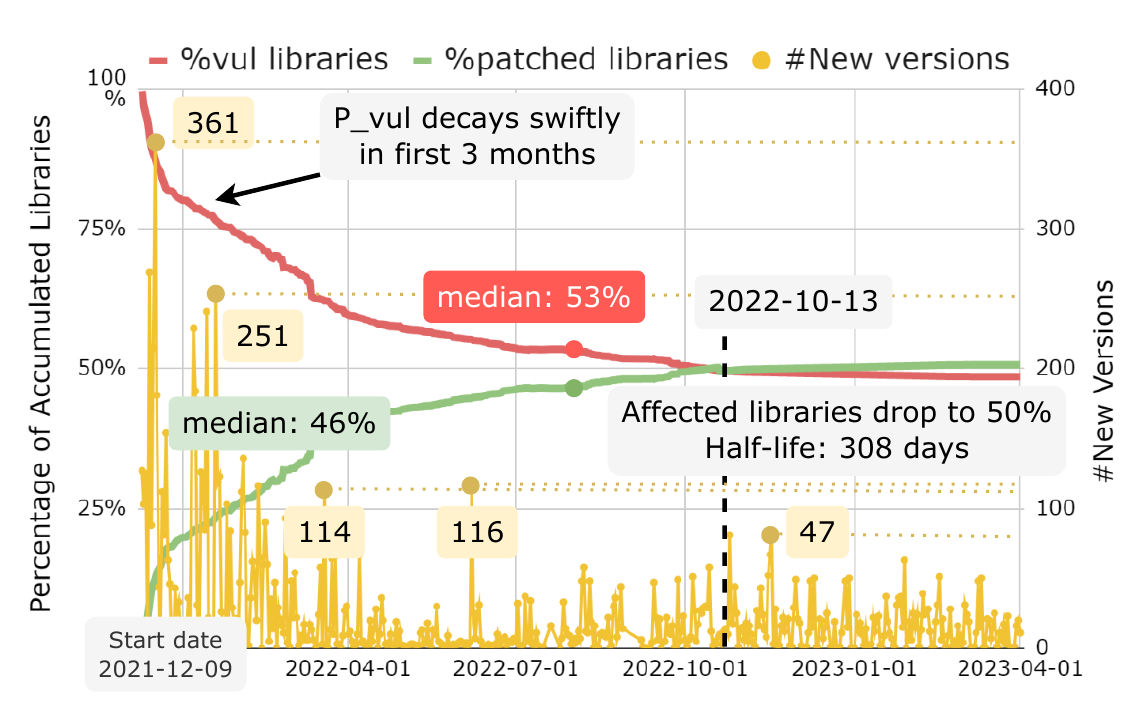}
  \caption{Accumulated Affected and Patched Libraries for Log4Shell} 
  \label{fig:vulpatch}
\end{figure}

In Figure~\ref{fig:vulpatch}, $P_{vul}$ and $P_{patch}$ are depicted by days. The sum of $P_{vul}$ and $P_{patch}$ is nearly $100\%$ because the number of the third category, \textbf{removed}, is negligible. The $P_{vul}$ reached $50\%$ in Oct 2022 and decayed much slowlier than before. Since $P_{vul}$ decays in a decelerating manner, Log4Shell would remain persistent in the ecosystem without abating for a long time. Hence, we define a metric, \textbf{Half-life}, to measure the time that $P_{vul}$ decays to $50\%$ from its initial value. The Half-life of Log4Shell can be measured based on days as $308$ days.

Although the $P_{vul}$ decays slowly, the number of newly released affected versions  decreases more quickly than $P_{vul}$ as in the same figure at the right axis. 
The number of new versions gradually decreases from the peak of $361$ when Log4Shell was initially exposed. 
It is seen that there were still new affected versions published after 15 months of exposure. We further investigated the depths of these versions and found out that $94\%$ of them were not first-level dependents. 
It suggests that the upstream dependencies of these affected libraries failed to upgrade \textit{log4j-core} in time.
Note that the number of new vulnerable versions has been fluctuating because the numbers are usually small on weekends.

To assess the prevalence of Log4Shell in real-world Java projects, we searched for vulnerable versions of \textit{log4j-core} in the dependency trees of the $9,220$ Maven projects we collected earlier. Our search revealed that $973$ ($10.55\%$) of these repositories had used \textit{log4j-core} in their dependency trees, out of which $392$ ($40.28\%$) were using the vulnerable versions of \textit{log4j-core}. We confirmed that none of these repositories had published vulnerable versions to MCR, which indicates that, besides libraries, end users were still using vulnerable \textit{log4j-core} versions in their projects after 15 months of disclosure.

\begin{tcolorbox}[size=title,opacityfill=0.2,breakable,boxsep=1mm]
\textbf{Finding 1:} The $P_{vul}$ of \textit{log4j-core} decayed rapidly to $65\%$ in the first 3 months upon disclosure. However, the decaying was decelerating, and it took $308$ days to reach its \textit{Half-life}. Log4Shell was still affecting $392$ GitHub Maven projects after 15 months.
\end{tcolorbox}

\subsubsection{Other Java Vulnerability Analysis}
\begin{figure}[!t]
  \centering \includegraphics[width=1\linewidth]{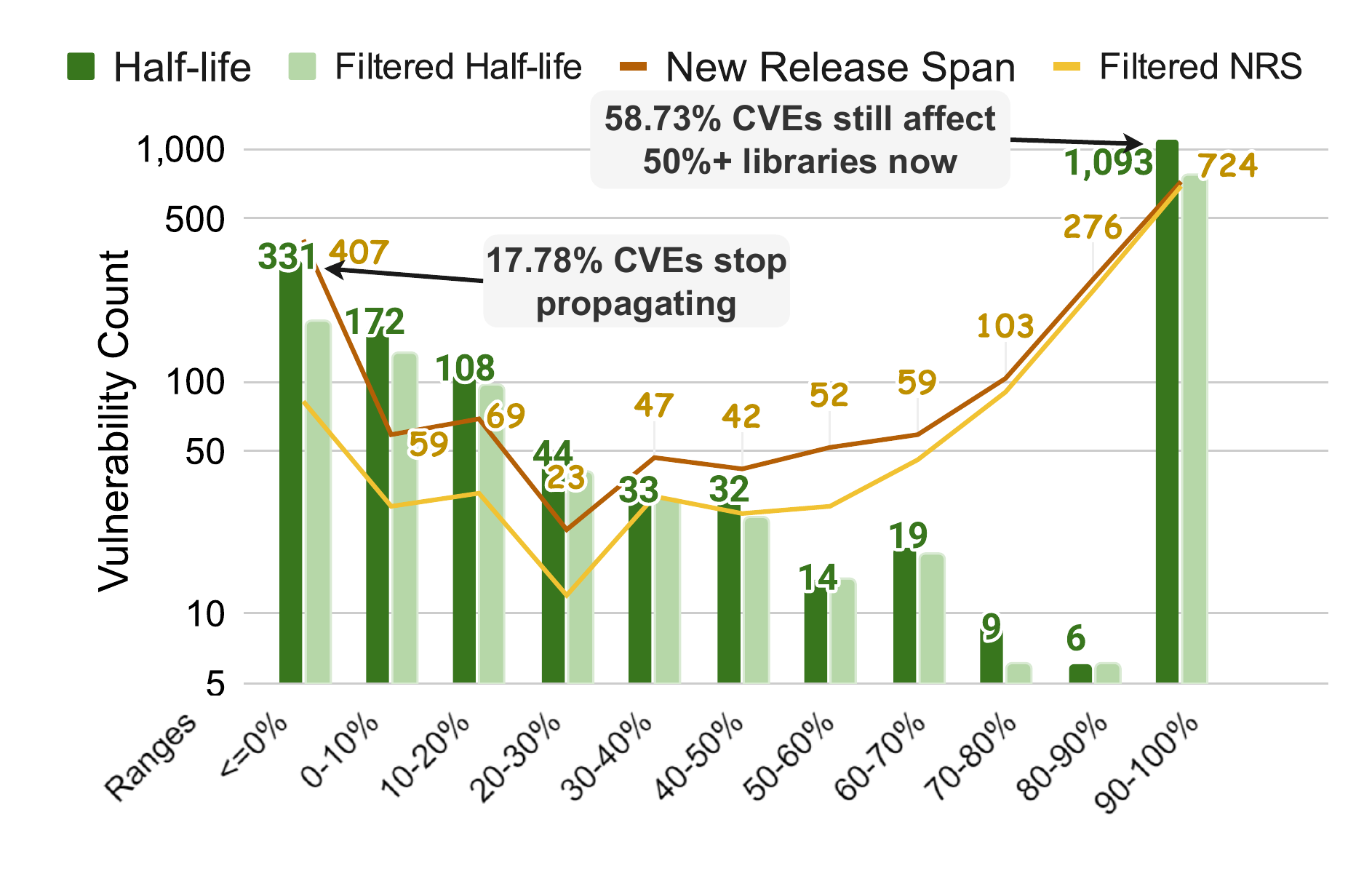}
  \caption{Distributions of Normalized Half-lives and New Release Span} 
  \label{fig:span}
\end{figure}

To find out if the decelerating decaying of $P_{vul}$ is prevalent for other vulnerabilities, we measured the $P_{vul}$ for all collected Java vulnerabilities as illustrated in Figure~\ref{fig:span}. Based on $P_{vul}$, the \textit{Half-lives} of vulnerabilities were derived. Since the exposure duration (from publishing date to data collection date 01 Apr 2023) varies greatly among vulnerabilities, we normalized \textit{Half-life} by dividing the exposure duration. While the \textbf{New Release Span} (NRS) was calculated by days from the CVE publishing date to the last date that affected versions are released. \textit{New Release Span} was also normalized by the same exposure duration per vulnerability. Because the number of affected libraries and versions vary greatly among vulnerabilities, the vulnerabilities with exceptionally few affected versions could bring deviations to the distributions. To ensure the representativeness of the vulnerability data set, we filtered out vulnerabilities that affected fewer than 100 versions and plotted the same normalized distributions in Figure~\ref{fig:span} as \textit{Filtered}. The number of filtered vulnerabilities was $1,319$.

The normalized Half-lives can be negative if the $P_{vul}$ already drops below $50\%$ before the vulnerability is published. Also, the normalized Half-lives can be $100\%$ if the $P_{vul}$ is still above $50\%$ by the data collection date. According to Figure~\ref{fig:span}, only $17.78\%$ of vulnerabilities have their $P_{vul}$ dropped below $50\%$ before the publishing of vulnerabilities, which means the rest $82.22\%$ of vulnerabilities affect $50\%+$ downstream libraries when they were disclosed. Even by the data collection date, $58.73\%$ of vulnerabilities still maintain over $50\%$ $P_{vul}$. Hence, it is concluded that most vulnerabilities persist and continue to affect downstream libraries, as seen in the case of Log4Shell.
In Figure~\ref{fig:span}, \textit{Filtered half-life}, denoted by light green bars, exhibits the distribution of filtered vulnerabilities. 
Because the numbers of filtered and pre-filtered vulnerabilities in most intervals are close to each other, it means vulnerabilities that were filtered out did not cause deviations. It is noteworthy that both ends of the distribution are higher than those in between, which means most vulnerabilities either were quickly remediated by downstream libraries or persisted in the ecosystem.

The normalized \textit{New Release Span} is used to indicate the impact of vulnerabilities in new releases.
In Figure~\ref{fig:span}, normalized \textit{New Release Span} is depicted by yellow lines. $39\%$ of vulnerabilities still have new affected versions released in the month of data collection (normalized \textit{New Release Span} is $100\%$), which indicates that nowadays there are still a non-trivial number of downstream developers who fail to upgrade their vulnerable dependencies. 
Although the distribution of \textit{New Release Span} is dissimilar to the half-life, they both have valley-like shapes, which proves the polarization in vulnerability remediation of the Maven ecosystem.
We further measured \textbf{Full-life} (the number of days that $P_{vul}$ drops to zero) instead of \textit{Half-life}, and it turned out that only 196 ($9.08\%$) of vulnerabilities have finite \textit{Full-lives}, which means only $9\%$ have all downstream libraries' latest versions fixed regardless of how long time it took.

\begin{tcolorbox}[size=title,opacityfill=0.2,breakable,boxsep=1mm]
\textbf{Finding 2:} $82.22\%$ of vulnerabilities affected $50\%+$ downstream libraries when they were disclosed. The \textit{vul rates} of vulnerabilities have been decaying in a decelerating manner over time, but till our data collection date, $58.73\%$ of them still maintained $50\%+$ \textit{vul rates}.
There are $39\%$ of vulnerabilities that still affect the new versions of downstream libraries that were released in the month of data collection. Only $9\%$ of vulnerabilities terminated their persistence.
\end{tcolorbox}

\subsection{RQ2: Study of Underlying Causes}
We aim to uncover underlying causes in this RQ. Inspired by the fact that $94\%$ of new versions are transitively affected by other vulnerable downstream libraries, we attempted to investigate the causes based on vulnerability propagation paths.
\subsubsection{Distribution of the causes}
\begin{figure*}[!t]
  \centering \includegraphics[width=1\linewidth]{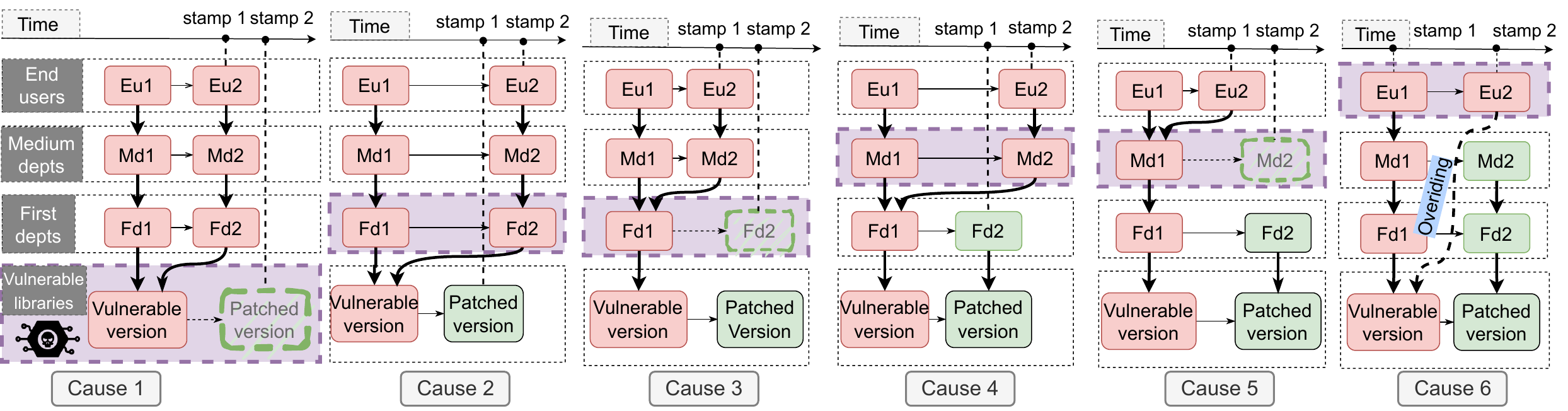}
  \caption{Scenarios of Different Causes} 
  \label{fig:scn}
\end{figure*}

We used a general model that included roles from the source of vulnerability to end users in the vulnerability propagation path as depicted on the left in Figure~\ref{fig:scn}. The roles are \textit{\textbf{Vulnerable libraries}}, \textit{\textbf{Medium dependents}}, and \textit{\textbf{End users}}. From RQ1, it is known that the misbehavior of these roles may block the patches from downstream libraries, which leads to persistent vulnerabilities. Hence, we further investigated what kind of misbehavior blocked the patches.
To clearly clarify causes without overlapping, the blockage of patches ascribes to the first role that conducts misbehavior during a bottom-up investigation because downstream roles automatically inherit configurations from the upstream.  

Based on the sequence of release time of two parties on each dependency relationship, we could summarize 6 causes out of three types of dependency relationships among 4 roles as illustrated in Figure \ref{fig:scn}.
In the figure, the rectangular box with dashed lines refers to the absent version vertex that is supposed to be present. And the boxes filled with purple color refer to the roles that are to blame for the patch blockage. For example, in the first column, the \textbf{Cause 1} is presented: The downstream dependents are affected by the vulnerability because the vulnerable library fails to release the patched version in time. Note that the \textit{\textbf{First Depts}} are split apart from \textit{\textbf{Medium Depts}} as an independent role because the first-level dependents directly determine the versions of the vulnerable libraries for the other medium dependents and can explicitly select the strategy between version ranges and \softs.

In Figure \ref{fig:scn}, \textbf{Cause 2} refers to the \textit{First Depts} still using the vulnerable versions even if the patched versions are available.
The \textbf{Cause 3} refers to the \textit{First Depts} failing to release new versions that depend on the patched versions so that the downstream dependents are forced to use the non-patched versions of \textit{First Depts}. Similarly, \textbf{Cause 4} and \textbf{Cause 5} refer to the corresponding misbehavior of \textit{Medium Depts}. The last \textbf{Cause 6} stands because the versions explicitly overridden by \textit{End users} are still vulnerable versions.

\begin{figure}[!t]
  \centering \includegraphics[width=1\linewidth]{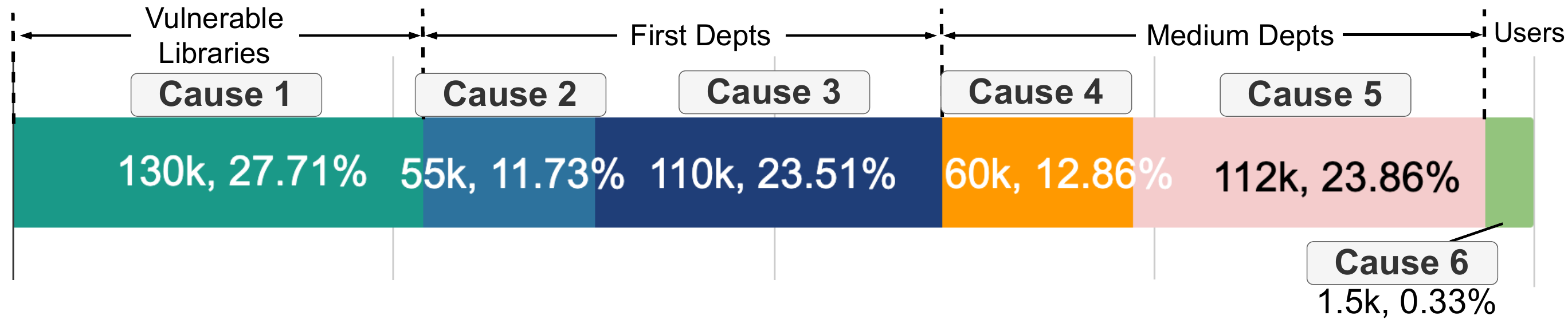}
  \caption{Proportions of Each Cause} 
  \label{fig:prop}
\end{figure}

Next, with the causes summarised, the importance of each cause is embodied by its proportion of occurrences. To avoid duplicated counts, we only counted the number of paths where blockage of patches occurred.  
Figure \ref{fig:prop} illustrates the proportions of each cause over all valid paths. Firstly, the \textit{Cause 1} is ruled out, because it is caused by the absence of patches instead of the blocked patches. 
Secondly, it is seen that the \textit{Medium Depts} account for most of the paths, which is $36.72\%$, and \textit{First Depts} account for a very close number of paths, which is $35.24\%$. Considering that the misbehavior of any role along the path could lead to the blocked patches, it is remarkable that \textit{First Depts} could affect the similar amount ($165k$ and $172k$) of paths as the rest all \textit{Medium Depts} at 2-15 depths, which proves that \textit{First Depts} is the most critical role regarding facilitating the patch adoption than other roles. Finally, 
although the proportion of \textit{Cause 6} is small, it proves that \textit{End users'} decisions are not always reliable. In fact, much domain knowledge and manual efforts are required for \textit{End users} to select the best version against all vulnerabilities.

\begin{tcolorbox}[size=title,opacityfill=0.2,breakable,boxsep=1mm]
\textbf{Finding 3:} It is concluded that misbehavior by \textit{First Depts} ($35.24\%$) and \textit{Medium Depts} ($36.72\%$) are guilty of the majority of affected paths. The \textit{First Depts} are the most significant role in terms of unblocking patches.
\end{tcolorbox}

In reality, most developers are only concerned by the vulnerabilities in their direct dependencies according to a study \cite{pashchenko2018vulnerable}. If a \textit{First Dept} uses the vulnerable version, the downstream libraries would automatically inherit the vulnerable version, which means that developers of \textit{First Depts} should be aware of the vulnerability and promptly upgrade vulnerable direct dependencies to patched versions instead of relying on downstream developers. Unfortunately, due to widely used \softs ($99\%$) in Maven, the versions specified for vulnerable libraries offer limited flexibility to upgrade against vulnerabilities. It would be unrealistic to force developers in Maven to swerve to version ranges abruptly because SemVer is not properly complied with in Maven and the backward compatibility has to be manually assured. Thus, to avoid reliance on developers, an automatic and scalable way to introduce flexibility to dependency versioning is required to mitigate persistent vulnerabilities.

\begin{tcolorbox}[size=title,opacityfill=0.2,breakable,boxsep=1mm]
\textbf{Finding 4:} The root cause of the misbehavior is the widely used \softs which greatly limit the flexibility of dependency version selection. Without flexible version ranges, the downstream libraries and applications are automatically prone to vulnerable dependencies even if patched versions are released. 
\end{tcolorbox}

\section{Methodology and Evaluation}
Because the limited flexibility hinders the spread of patches, we aim to introduce the flexibility to unblock the patches. First, we reviewed the existing solutions to identify the pros and cons, based on which, our solution \tool is proposed and evaluated to answer the following research questions:

\noindent $\bullet$ \textbf{RQ3}: \textbf{\textit{How do existing solutions address persistent vulnerabilities?}}

\noindent $\bullet$ \textbf{RQ4}: \textbf{\textit{How effective is \tool regarding mitigating the persistence of vulnerabilities?}}

\subsection{RQ3: Review of Existing Solutions}
The solution recommended by Maven is the semantic version ranges \cite{mavenranges}. Unfortunately, considering that SemVer is not properly complied with, instability could be introduced by ranges so that ranges are rarely adopted. Apart from ranges, the most used approach is the transitive version override.
If the versions of transitive dependencies are vulnerable, any dependent can override the transitive vulnerable versions by \textit{dependencyManagement}. Hence, as solutions supported by Maven, ranges and version overriding can be used to mitigate the persistence of vulnerabilities. 
Note that besides Maven, other popular Java Package Managers, Gradle~\cite{gradledep} and Ivy~\cite{ivydep} implement similar overriding mechanisms, \textit{Dependency Constraints} and \textit{Dependency Overriding} respectively to determine the versions of transitive dependencies if the transitive dependencies exist. Thus, we refer to this overriding mechanism as \textit{\depv Overriding}.
There are also other workarounds, such as tampering with local libraries of vulnerable dependencies to manually backport patches for deployment environments. But temporary workarounds are too infrequently used to be discussed. Furthermore, \textit{exclusion} supported by Maven is not discussed either, because it is used to exclude unused transitive dependencies which are not worth mitigating the vulnerabilities for.

\subsubsection{Study of the Usage of Ranges}
From the $82m$ collected dependency relationships, only $637,783$ ($1.02\%$) of them use the semantic version ranges. Out of the range-use dependencies relationships, $29,556$ ($4.63\%$) are specified for the vulnerable libraries by \textit{First Depts}. We further investigated these vulnerability-related ranges to reveal how many vulnerabilities can be automatically bypassed.

\begin{figure}[!t]
  \centering \includegraphics[width=1\linewidth]{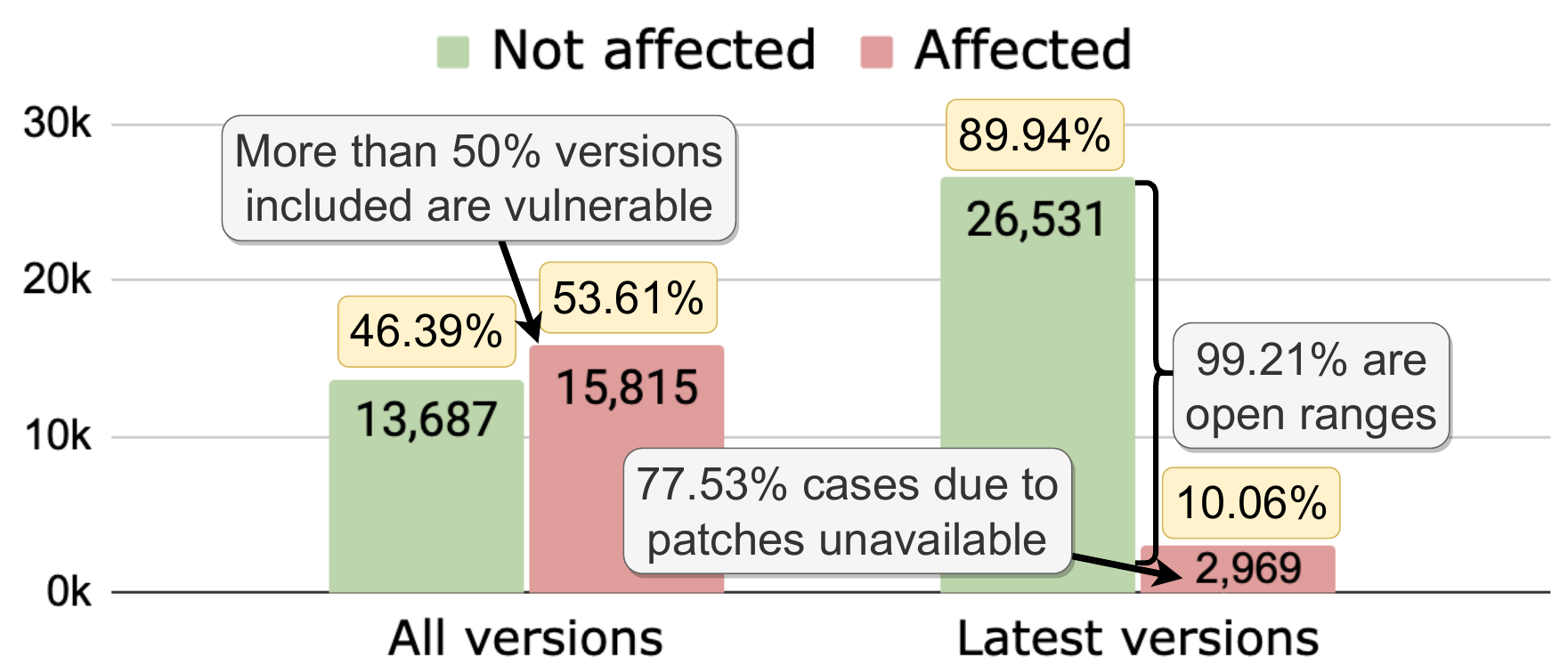}
  \caption{Usage of Vulnerability Related Version Ranges}
  \label{fig:range}
\end{figure}

As illustrated by the left two bars in Figure~\ref{fig:range}, considering all versions within the ranges, $53.61\%$ of versions are vulnerable. However, Maven would usually select the latest (semantically highest) version in a version range as the resolved version of the dependency. Thus, if only the latest versions in these ranges are considered, the proportion of vulnerable latest versions drops to $10.06\%$ in the right 2 bars in Figure~\ref{fig:range}, which proves that version ranges can effectively free dependents from vulnerabilities if patched versions are included.
To understand how patched versions were introduced, we went through the ranges whose latest versions are not vulnerable. Out of the $26,531$ non-vulnerable versions, $99.21\%$ of them belong to ranges that are actually right open ranges, such as \textit{[1.1,)} without defining the upper bounds. Because the right open ranges would always be resolved to the latest version, the potential breaking changes could be introduced to the dependents whenever any incompatible new versions are released.

\begin{tcolorbox}[size=title,opacityfill=0.2,breakable,boxsep=1mm]
\textbf{Finding 5:} Although the version ranges allow flexible upgrades of vulnerable dependencies, they are rarely used in Maven ($1.02\%$). The fact that the latest versions of $89.94\%$ of version ranges of vulnerable libraries were no longer vulnerable proves the effectiveness of version ranges.
However, $99.21\%$ of ranges that successfully bypassed vulnerabilities were open ranges that are subject to unpredictable incompatibility issues. Hence, to properly use version ranges, compatibility has to be assured.
\end{tcolorbox}

\subsubsection{Study of the Version Overidding}
Because only \textit{Medium Depts} and \textit{End users}, the indirect dependents of vulnerable libraries, would use \textit{\depv Overriding} to control the versions of transitive dependencies,
we study the effectiveness of \textit{\depv Overriding} for them. In total, there are $639,710$ ($6.50\%$) POM files for versions that use \textit{dependencyManagement}, from which we extracted the overridden versions per POM file. 
After matching the overridden versions with vulnerability mappings, we found $295,951$ POM files have overridden versions of vulnerable libraries.
Then, these files were categorized into 2 cases in Table~\ref{tab:dm} regarding whether they bypassed the vulnerabilities: (1) \textit{Affected}: Any overridden version in the POM file was still vulnerable. (2) \textit{Bypass}: 
The default version was vulnerable but the overridden was not. 
Note that there was overlapping because a POM file may have multiple overridden versions. 

It turned out $86\%$ of POMs bypassed the vulnerabilities in transitive dependencies because probably their developers were aware of the vulnerabilities and explicitly addressed them with \textit{dependencyManagement}. However, it is surprising that $72\%$ ($214,933$) POM files both bypassed some CVEs and introduced other CVEs at the same time. Only $14\%$ of POMs completely bypassed all vulnerabilities. It implies that fixing vulnerabilities with version overriding is a non-trivial job, which is the first weakness of version overriding, \textbf{Knowledge and efforts}. The developers must equip with extensive domain knowledge of vulnerabilities and invest efforts to ensure their eradication. 
Although version overriding is able to address vulnerabilities for the current project within a POM file, it is not inheritable according to Maven Specification \cite{mavenver} so that it does not benefit the downstream libraries. Another weakness of version overriding is \textbf{Non-inheritability}, because of which,  
Since the version overriding only works for current projects instead of dependents that depend on the projects, the vulnerable versions are still being used by downstream libraries unless all developers along the propagation path conduct the same overriding. Therefore, 
the patch versions cannot be automatically adopted by downstream users. 
In conclusion, the version overriding can only serve as a temporary workaround instead of boosting the self-healing of the ecosystem.

\begin{table}[!t]
\footnotesize
\setlength{\tabcolsep}{4pt}
\caption{Counts of POMs with \textit{dependencyManagement}}
\begin{tabular}{rrrr}
\toprule
              \textbf{With vul libraries}           & \textbf{Affected by CVEs}    & \textbf{Bypass CVEs}   & \textbf{Overlapping}         \\
 \midrule
 \multicolumn{1}{r}{295,951} & \multicolumn{1}{r}{254,043 (86\%)} & 256,841 (87\%) & $214,933$ (72\%)\\
\bottomrule
\end{tabular}
\label{tab:dm}
\end{table}

\begin{tcolorbox}[size=title,opacityfill=0.2,breakable,boxsep=1mm]
\textbf{Finding 6:} The adoption rate of dependency version overriding is $6.50\%$ and only $14\%$ of adopters completely bypassed all vulnerabilities. Because dependency version overriding requires knowledge and manual effort and is unable to benefit downstream users due to non-inheritability, it is not effective in eliminating persistent vulnerabilities.
\end{tcolorbox}

Another solution worth discussing is Plumber~\cite{wang2023plumber} which addresses the persistent vulnerabilities in the NPM ecosystem. Plumber employs a dependency graph to identify the dependents that block fixes of vulnerabilities. Subsequently, it endeavors to upgrade the blocking dependents to compatible versions. If upgrading is not possible within the bounds of compatibility, Plumber generates remediation suggestions, such as backporting and migration, both of which require manual intervention. However, Plumber is not applicable to Maven, because it relies on compatible ranges that are pre-specified by developers, a feature that is prevalent in NPM~\cite{liu2022demystifying} but not in Maven, which further necessitates the compatible version ranges for Maven.

\subsection{RQ4: Methodology and Evaluation of \tool}
\subsubsection{Requirements of the Solution}

Based on the previous research question, existing solutions, such that open version ranges are subject to breaking changes and \depv overriding is non-inheritable and requires intensive manual efforts. Despite the limited usage, version ranges were proven to be effective for unblocking the patches. However, due to legacy reasons, developers predominantly utilize \softs, making it impractical to mandate a shift toward version ranges, not to mention that version ranges have to be manually curated by developers. 
Therefore, our objective is to propose an automated solution for restoring version ranges of both vulnerable libraries and dependencies that transitively depend on vulnerable libraries from \softs. By restoring the version ranges, vulnerability fixes within the ranges can propagate smoothly and automatically to downstream users. 
Moreover, for the purpose of ecosystem-level implementation, \tool should possess the ability to continuously monitor the Maven ecosystem for blocking dependents and promptly provide the restored version ranges along with corresponding suggestions to the developers of such blocking dependents. This approach would expedite the propagation of patches throughout the ecosystem.

\subsubsection{{Design of \tool}}
\label{sec:checker}

To this end, we have proposed \tool, which comprises a server-side edition and a client plug-in. The client plug-in for \tool can be integrated into a developer's workflow as a Maven plug-in. For a Maven project, this plug-in can automatically replace the \softs in the POM file with curated compatible version ranges, and the developer can effortlessly publish the updated POM file with version ranges to benefit downstream users. On the other hand, the server-side edition of \tool employs the \searchalg algorithm to continuously monitor an up-to-date dependency graph for instances of vulnerability fix blockage caused by \softs. When a blockage is detected, \tool calculates compatible version ranges for the vulnerable constraints and reports this suggestion of version ranges to the relevant developers. 

\begin{figure}[!t]
  \centering \includegraphics[width=1\linewidth]{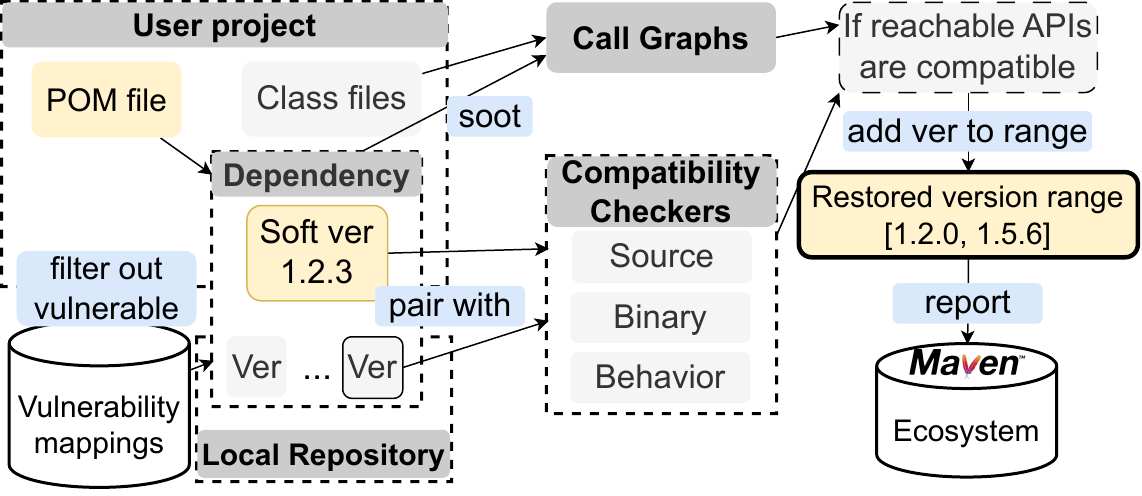}
  \caption{Overview of \tool} 
  \label{fig:demo}
\end{figure}

As depicted in Figure \ref{fig:demo}, we first introduce the plug-in that accepts a dependency with \soft and class files of the project as input. Given that version ranges specified by developers typically consider compatibility and functionality, \tool aims to ensure them for the restored version ranges. Specifically, given a \soft $v_{s}$, \tool retrieves sorted candidate versions $V_{cand}= \{v_1, v_2, ..., v_n\}$ from the MCR as a list as well as the version and vulnerability mappings from the dependency graph. Then \tool determines which versions from $V_{cand}$ should be included in the restored range $V_{r}$ to ensure $V_r$ is more secure, flexible, and compatible. We further formulate the problem into a Multi-Objective Optimization problem:
\begin{itemize}[leftmargin=9pt]
    \item \textbf{Objective 1} (Primary): The maximum number of vulnerabilities for all versions in $V_r$ is minimized to guarantee any version resolved by Maven is more secure than $v_s$.
    \item \textbf{Objective 2} (Secondary): $V_r$ should include as many candidate $v$ as possible for better flexibility.
    \item \textbf{Constraint 1}: $V_r$ must be compatible with $v_{s}$.
    \item \textbf{Constraint 2}: any $v_r$ in $V_r$ must has not greater vulnerabilities than $v_{s}$ to ensure the effectiveness of restoration.
\end{itemize}

\scalebox{0.95}{
\begin{minipage}{\linewidth}
    \begin{equation}
    \begin{aligned}
     \footnotesize
        min\quad &f_1 = max(\sum_{n=1}^{dt(v_r)} count_{vul}(n))\\[0pt]
        max\quad &f_2 = |V_r|\\[1pt]
        s.t.\quad &c_1: compatibility(v_{s}, v_1) = 1 |\forall v_r \in V_r\\[0pt]
               &c_2: \sum_{n=1}^{dt(v_r)}count_{vul}(v_r)\leq \sum_{n=1}^{dt(v_{s})}count_{vul}(v_{s})| \forall v_r \in V_r\\[0pt]\nonumber
    \end{aligned}
    \end{equation}
\end{minipage}
}
where $dt(v) = \{n_1, n_2, ..., n_t\}$ is to resolve a dependency tree from the version $v$. the total vulnerabilities of $v$ are the sum of numbers of vulnerabilities associated with each node in $dt(v)$ to include transitive vulnerabilities.

\begin{algorithm2e}[t!]
    \small
 \setcounter{AlgoLine}{0}
 \caption{Algorithm of \tool}
 \label{alg:1}
 \DontPrintSemicolon
 \SetCommentSty{mycommfont}
 {
     \KwIn{\soft $v_{s}$, candidate versions $V_{cand}$, class files $f$}
     \KwOut{Restored version range $V_r$}
     $V' \gets set(v_s)$\;
     $dt_{v_s} \gets dependencyTree(v_s)$ \;
     $vul_{v_s} \gets queryCVE(dt_{v_s})$ \;
     \ForEach{$v_{cand}$ in $V_{cand}$}{
        $dt \gets dependencyTree(v_{cand})$\;
        $vul \gets queryCVE(dt)$\;
        \If{$\And vul\leq vul_{v_s}$}{
            $V' \gets v_{cand}$\;
        }
        $sort(V')$\;
        $V_{upper} \gets V'.truncate(v_s, v_{last})$\;
        $V_{lower} \gets V'.truncate(v_{first}, v_{v_s})$\;
        \ForEach{$v$ in $ V_{upper}$}{
            $api = compatibilityCheck(v_s, v)$ \;
            \If{$\neg reachable(f, api)$}{
                $V_r \gets v$\;
            }
        }
        \ForEach{$v$ in $ reverse(V_{lower})$}{
            $api = compatibilityCheck(v_s, v)$ \;
            \If{$\neg reachable(f, api)$}{
                $V_r \gets v$\;
            }
        }
    }
    \ForEach{$v_r$ in $V_r$}{
        \If{$queryCVE(dt(v_r))> min(Vul(V_s))$}{
            $V_r \; \textbf{remove}\; v_r$\;
        }
    }
    \ForEach{$v_r$ in $V_r$}{
        \If{$\neg unitTest(v_r)$}{
            $V_r \; \textbf{remove}\; v_r$\;
        }
    }
    \
    \KwRet{$V_r$}
}
\end{algorithm2e}

We implemented Algorithm~\ref{alg:1} in \tool to solve the problem above. 
From L1-L8, \tool first queries the number of vulnerabilities for the resolved dependency tree of \soft $v_s$ and each version in $V_{cand}$. To adhere to constraint $c_2$, the versions with more vulnerabilities than $v_s$ are filtered out. Then from L9-L11, \tool sort the filtered versions in a SemVer order and split the list of versions into the upper and lower parts by $v_s$ to add potential versions bi-directionally for more candidates. For both the upper and lower parts, \tool checks the compatibility between candidates and $v_s$. Note that the compatibility checkers employed by \tool can handle all types of code-based compatibility as specified in the Oracle documentation~\cite{oraclecomp}, including Source, Binary, and Behavioral Compatibility. The Source and Binary Compatibility are ensured by two commonly used tools, revapi and jcp~\cite{revapi, japi-compliance-checker} with high accuracy. For Behavioral Compatibility, we used the only static detector Sembid~\cite{zhang2022has}. 
Specifically, \tool calculates incompatible APIs with the checkers and compares them with the reachable APIs collected from the call graphs. The call graphs are constructed with Soot Spark~\cite{sootspark} from the class files of the project and byte code of the dependency to determine whether any incompatible API is reachable. If a candidate version has no reachable incompatible APIs, it is included in the range. 
In L13, compatibility checkers serve as a pre-filter for the final validation because they are static and more efficient than testing.
In L20-L22, \tool excludes versions that have more vulnerabilities than the minimum required in order to satisfy Objective $f_1$. In L23-25, given the heavier resource demands of unit tests, \tool further excludes versions failing the unit test serving as the final validation.

Regarding the server-side edition of \tool, the initial step involves identifying the blocking dependents, denoted as \textit{First Depts}, by means of the \searchalg algorithm. The plug-in running on the server proceeds to calculate and test the compatible version ranges using the repositories stored in our database. If a version range covering the patched version is successfully restored, \tool generates a report that is sent to the relevant developer. In cases where range restoration fails, \tool attempts to locate the \textit{Second Dept} of the failed \textit{First Dept} from the dependency graph and calculates the restorable range towards the \textit{First Dept} instead of the vulnerable library. This is because \textit{First Dept} is the direct dependency of \textit{Second Dept} and only specified versions of direct dependencies are transitive for the rest of the dependents. This process is repeated 10 times until no range can be restored.

\subsubsection{Evaluation of \tool}
\label{sec:evaluationoftool}
To showcase the effectiveness of \tool in real-world scenarios, we initially evaluated the plug-in on a dataset of $252$ GitHub repositories that included vulnerable versions of \textit{log4j-core} in their dependency trees, as of 01 Apr 2023. Subsequently, we conducted a large-scale evaluation on another dataset to demonstrate the effectiveness of \tool for mitigating persistent vulnerabilities in the Maven ecosystem.

\noindent $\bullet$ \textbf{Evaluation of Plug-in:} 
\textbf{Dataset:} 
From the $9,220$ repositories in Section~\ref{sec:evaluation}, we retrieved the dependency by Maven command and check if any vulnerable log4j version was still in use. $374$ repositories were derived. Only $252$ of them could be successfully compiled and tested. \textbf{Results:} We ran \tool to only restore the version ranges for \textit{log4j-core} in $252$ parent POM files to control the variables. $160$ secure ranges of them were restored with a $63.49\%$ restoration rate. Before running the compilation and unit tests, $179$ raw ranges were statically calculated, which means unit tests could reduce $19$ false positives. It should be noted that false negative cases were present in our evaluation, as false alerts produced by static compatibility checkers exclude the potential candidate versions, and unit tests only narrow the range but not widen it. Thus we manually checked the failed cases and found that $11$ could have been restored. 
The results were summarized in Table~\ref{tab:demo} with the actually restorable repositories accounting for $67.85\%$. Although \tool had some false negatives, it hardly introduced false positives that could break current and downstream projects. It was proven that $93.57\%$ of restorable version ranges could be automatically restored by \tool.

\begin{table}[!t]
\setlength{\tabcolsep}{4pt}
\footnotesize
\caption{Results of \tool}
\scalebox{1}{
\begin{tabular}{lrrrrr}
\toprule
& \multicolumn{1}{l}{\textbf{Restored}} &  \multicolumn{1}{l}{\textbf{Failed Unit Tests}} & \textbf{Restore Rate}               & \textbf{Recall/Precision} \\
\midrule
*GT & 171                                                    & N.A.   & \multicolumn{1}{r}{67.85\%} & N.A.            \\
\tool    & 160                                                    & 19                            & 63.49\%                         & 93.57\%/100.00\% \\
\bottomrule
\end{tabular}}
\label{tab:demo}
\begin{enumerate}
\item \footnotesize{GT stands for ground truth.}
\end{enumerate}
\end{table}
\begin{figure}[!t]
  \centering \includegraphics[width=1\linewidth]{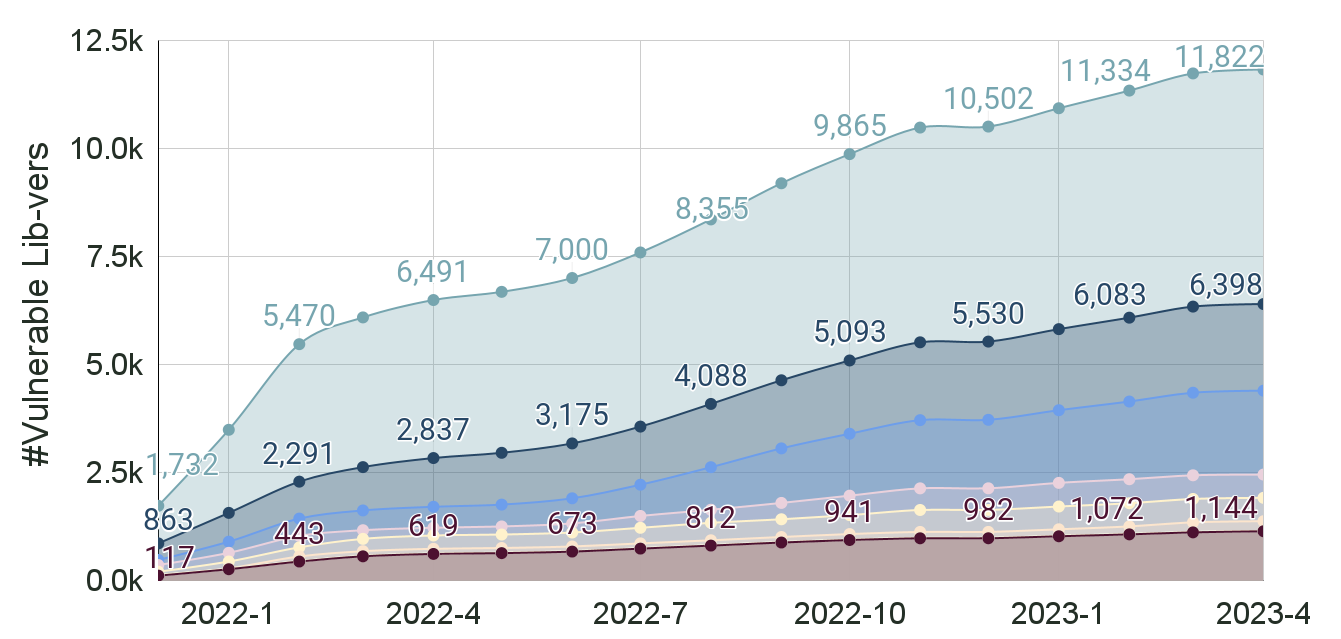}
  \caption{Number of Vulnerable Lib-vers over Months after Applying \tool to Dependents at 1-10 Depths} 
  \label{fig:tool}
\end{figure}

\noindent $\bullet$ \textbf{Evaluation of Server-side Edition:} 
\textbf{Dataset:} 
To simulate a scenario where developers of downstream dependents adopt the version ranges generated by \tool upon the disclosure of a vulnerability, we compiled a list of all affected libraries and versions published after the disclosure of Log4Shell. This resulted in a total of 11,822 library-version pairs, denoted as lib-vers. 
\textbf{Results:} Regarding the lib-vers, \tool first restored ranges for the \textit{First Depts} at Depth 1. This resulted in a successful restoration of $486$ out of $668$ dependent lib-vers. The generated ranges were then applied to the dependency graph, and we performed \searchalg again to retrieve the affected \textit{Second Depts} at Depth 2 on an updated graph. Out of $927$ \textit{Second Depts}, $731$ were successfully restored. We repeated this process for a total of 10 depths and evaluated the number of vulnerable lib-vers over time, as shown in Figure~\ref{fig:tool}.

In total, it took $4,110$ iterations to successfully restore $3,109$ version ranges with a $75.64\%$ restoration rate. As a result, $90.32\%$ of the vulnerable lib-vers were successfully remediated from Log4Shell, leaving only $1,144$. 
It is clear that the number of vulnerable lib-vers increases much more slowly over time after applying \tool to the 10th dependents than the primitive state. This suggests that the propagation of the vulnerability was effectively suppressed from the beginning upon disclosure of Log4Shell. Moreover, it is observed that the 
number drops $45.95\%$ when \tool is only applied to the \textit{First depts} at Depth 1, which indicates that \textit{First depts} have a significant impact on downstream libraries yet not enough to suppress the propagation. Also, the marginal effect of \tool drops fast as depth goes deep in the Figure and there were only 8 ranges to restore at Depth 9 and 10, which means Depth 10 is effective enough against persistent Log4Shell. 

However, there still remained $1,144$ unfixed lib-vers requiring manual intervention. We categorized the remaining cases into three: (1) \textbf{No compatible patched versions to upgrade} (481 cases, $42\%$): \tool found there was no version satisfying the constraints. For these cases, \tool generated a report with breaking APIs and call chains with suggestions of manual fixes for developers to resolve the incompatibility. (2) \textbf{No secure versions available} (592 cases, $52\%$): This mostly happens for dependents at Depth 2+ because their direct dependencies may not have published a secure version that transitively depends on a patched version of \textit{log4j-core}. It is a common case, especially for a newly disclosed vulnerability, for which, \tool would suggest the developers find a substitution if the vulnerable library is reachable. If not reachable, a suggestion to exclude the vulnerable library would be suggested. On the other hand, \tool will continue to monitor the availability of patched versions. (3) \textbf{Internal error} (71 cases, $6\%$): These were caused by issues irrelevant to the design of \tool, namely, failed jar downloading, failed call graph generation, and the errors of compatibility checkers.

\begin{tcolorbox}[size=title,opacityfill=0.2,breakable,boxsep=1mm]
\textbf{Finding 7:} 
Our evaluation demonstrated that \tool, as a plug-in, was successful in restoring secure version ranges for $63.49\%$ of the $252$ real-world GitHub repositories, with a high recall of $93.57\%$. In a simulated experiment, the server-side edition of \tool was able to restore version ranges for $3,109$ ($75.64\%$) of the dependents which successfully remediated $10,678$ ($90.32\%$) of downstream vulnerable projects. 
\end{tcolorbox}

\section{Discussion}

\noindent $\bullet$ \textbf{Compatibility check should be aligned with SemVer, especially in Maven.} Many studies~\cite{raemaekers2014semantic,raemaekers2017semantic,ochoa2021breaking,decan2019package,zhang2022has} have revealed that SemVer has not been well adhered to by developers in the Maven ecosystem, leading to prevalent \soft. Although \tool could restore version ranges to include patched versions, the patched versions must be those already published. To timely apply the patches upon releases, version ranges have to be open ranges or semi-open ranges, e.g. caret range $\string^1.2.3$~\cite{caretrange}, which requires strict compliance with SemVer to assure compatibility. 
Therefore, in the long run, \tool could mitigate the persistent vulnerabilities, but only the widely used and strictly backward compatible open version ranges could nip them in the bud.

\noindent $\bullet$ \textbf{More efforts and resources should be leaned on widely used but poorly maintained libraries.} 
As revealed by our evaluation in Section~\ref{sec:evaluationoftool}, there were $592$ cases without patched versions to upgrade to. Following a manual investigation, it was discovered that several libraries served as dependencies in a large number of projects or libraries, but their maintenance was inadequate. To address persistent vulnerabilities, it is imperative that the maintainers explicitly release patched versions for the benefit of downstream users. Therefore, this kind of libraries should arouse the collective awareness of the community, and the resources of open-source software governance should be directed towards these widely-used but poorly-maintained libraries to promote a more secure ecosystem.

\section{Threats of Validity}
\label{sec:threat}
The primary threat of the study is the assumption that dependents of vulnerable libraries were considered affected without fine-grained reachability or triggerability analysis. Because analyzing the reachability of all vulnerabilities in the entire Maven ecosystem at a large scale is quite expensive, we did not take it into consideration. Furthermore, vulnerable libraries are also packaged into the deployment environment, and having vulnerability is not a secure practice because they could be exploited someday given the evolving source code. Hence, to promote the best security practice in the Maven ecosystem, we made such an over-assumption.

Another threat is the assumption that successful compilation and passing unit tests after applying version ranges generated by \tool are sufficient to confirm successful version range restoration. However, in real-world software development, unit tests have limited coverage, and passing them does not necessarily guarantee that the restored version ranges satisfy all requirements of developers. Despite this limitation, unit tests are a critical component of deployment and are currently the most convenient validation approach available.

The last threat is the accuracy of the algorithm \searchalg that is used to track the downstream libraries. The first factor affecting the accuracy is the dependency graph sourced from MCR, and a few POM files in MCR could be unavailable leading to incomplete dependency edges in the graph. Another factor is that the environment requirements of dependencies were ignored, which could lead to false positives because some dependencies are only installed in certain environments, such as Windows. However, these factors were proven to be corner cases in the validation experiment in Section~\ref{sec:deptree} so that the overall conclusions are not undermined.

\section{Related Work}
\subsection{Persistence of Vulnerabilities}
Researchers~\cite{wu2023understanding, li2021pdgraph, pashchenko2020qualitative, wang2023plumber,benelallam2019maven} have evaluated the propagation of vulnerabilities within the Maven ecosystem and recognized the long-term persistence of some vulnerabilities. 
Developers tend to address reachable vulnerabilities more than unknown ones due to the potential for exploitation, as revealed by Wu et al.\cite{wu2023understanding}. Pashchenko et al.\cite{pashchenko2020qualitative} found that upgrades to vulnerable dependencies are often delayed due to potential breaking changes.
Li et al.\cite{li2021pdgraph} conducted a similar quantitative study using a dependency graph integrated with vulnerabilities. Benelallam et al.\cite{benelallam2019maven} proposed the Maven dependency graph that has been widely used for ecosystem vulnerability analysis.
Plumber~\cite{wang2023plumber} proposed by Wang et al. is a viable approach to address persistent vulnerabilities in NPM but not applicable to Maven because it relies on the pre-defined version ranges prevalent in NPM. Although insights have been highlighted, they have not proposed any tailored solution for persistent vulnerabilities for Maven.

\subsection{Remediation for Maven Vulnerabilities}
Regarding Maven vulnerability remediation, many solutions have been proposed~\cite{du2018refining,zhang2023compatible,mir2023effect,scorecard, sca,openssf,ossfwgbe,kula2015trusting,kulaamodeling,kula2014visualizing,massacci2021technical,mitropoulos2013dismal,alqahtani2016sv}. 
Coral~\cite{zhang2023compatible} is a systematic approach to address the vulnerabilities in the dependency trees of user projects. Du et al.~\cite{du2018refining} constructed a patch tracing system to locate patches to remediate the vulnerabilities.
Industrial organization, OpenSSF~\cite{openssf}, has proposed the best practice guidance~\cite{ossfwgbe} and a tool, Scorecard~\cite{scorecard}, for developers on managing vulnerabilities in dependencies for developers.
Software Composition Analysis (SCA)~\cite{sca,zhao2023fse,zhan2021atvhunter,zhan2021research,wu2023ossfp} tools have also been widely adopted to assist in the mitigation of vulnerabilities persistent in users' projects. 
However, these studies focused on user-oriented remediation by developers instead of ecosystem-wide vulnerability mitigation.

\subsection{Dependency Versioning in Modern Ecosystem}
Many
researchers have recognized the significance of the dependency versioning scheme for the security and stability of open-software ecosystem.~\cite{dietrich2019dependency,decan2019package,ochoa2021breaking,raemaekers2014semantic,osvinsight,wang2018dependency} 
Dietrich et al.~\cite{dietrich2019dependency} studied 17 package managers to investigate the dependency versioning recommended by them and found Maven heavily uses \soft leading to low flexibility of dependency versions. 
Google~\cite{osvinsight} released a blog about persistent vulnerabilities like Log4Shell and pinpointed that the \softs could be a cause of the persistent vulnerabilities. Decan et al.~\cite{decan2019package} reviewed the SemVer compliance in 4 popular package managers and summarized guidance for developers to better comply with SemVer. 
These works highlight the limitations of dependency versioning in modern ecosystems, including the lack of flexibility in dependency management within Maven. As a solution, we proposed \tool to restore the flexibility of version ranges within the Maven ecosystem.

\section{Conclusion}

In order to find a solution that addresses ecosystem-wide persistent vulnerabilities, 
we conducted an empirical study that revealed that $58.73\%$ of vulnerabilities still impacted more than $50\%$ of downstream
libraries in the Maven ecosystem nowadays. 
Through this study, we quantitatively substantiated that blocked patches caused the persistence of vulnerabilities. The existing solutions are either not scalable or subject to breaking changes. Hence, we proposed \tool as a scalable and automatic approach with compatibility assurance to unblock the vulnerability patches.  
Through evaluation, \tool achieved $93.57\%$ recall and restored $3,109$ ($75.64\%$) ranges, which remediated $10, 678$ ($90.32\%$) vulnerable downstream projects.

\noindent $\bullet$ \textbf{Data Availability.} The experiment data set and algorithm are available at our website~\cite{dataset}.

\section{Acknowledgement}
This study is supported under the RIE2020 Industry Alignment Fund – Industry Collaboration Projects (IAF-ICP) Funding Initiative, as well as cash and in-kind contribution from the industry partner(s). This research is partially supported by the National Research Foundation Singapore and DSO National Laboratories under the AI Singapore Programme (AISG Award No: AISG2-RP-2020-019), the NRF Investigatorship NRF-NRFI06-2020-0001,
the Ministry of Education, Singapore under its Academic Research Fund Tier 3 (MOET32020-0004). Any opinions, findings and conclusions or recommendations expressed in this material are those of the author(s) and do not reflect the views of the Ministry of Education, Singapore.
\bibliographystyle{IEEEtran}
\bibliography{acmart}
\end{document}